\documentstyle[eqsecnum,aps,prd]{revtex}
\begin{document}

\title{Spacetime at the Planck scale and Path Integral}

\author{G. Mangano}

\address{INFN, Sezione di Napoli, Mostra d'Oltremare Pad. 20, I-80125
Napoli, Italy\\E-mail: mangano@na.infn.it}

\maketitle
\begin{abstract}
{We discuss a path integral formalism to introduce noncommutative
generalizations of spacetime manifold in even dimensions, which have been
suggested to be reasonable effective pictures at very small length scales,
of the order of Planck length.}
\end{abstract}
\bigskip
{\small
\noindent {\it Amor, ch'a nullo amato amar perdona,\\
                          mi prese del costui piacer s\`{\i} forte,\\
                          che, come vedi, ancor non m'abbandona.\\

$~~~~~~~~~~~~~~~~~~~~~~~~~~~~~~~~~~~~~~~$To Carmen}\\}

\section{Introduction}

A fascinating open problem of modern physics is a satisfactory
understanding of the structure of spacetime at very short length scale, of
the order of the Planck length $\lambda_P \sim 10^{-33}~cm$. On one side
the difficulties of quantization of gravity in canonical ways can be only
due to technical reasons, and we have simply to wait for new powerful
mathematical tools. What is implicit in this possibility is however that
the very basic structure of spacetime is a customary manifold with
suitable differential and topological structures, which is the
arena for interactions among quantized fields. On the other hand, many
arguments on operational limits on position and time measurements have been
considered in literature \cite{a}, suggesting that this description should
be modified in a way which is reminiscent of the quantization of phase
space in ordinary quantum mechanics. It is worth stressing that, in this case,
it is the
very notion of {\it manifold} which now undergoes a dramatic change. If we
adopt a dual point of view, it is well known that the (commutative)
algebra of smooth functions over spacetime manifold
contains, using
Gelf'and--Naimark reconstruction theorem, all informations on the
underlying space. Switching on the Planck length then amounts to consider
the noncommutative algebra still generated by positions and time $X^\mu$,
now looked upon as generators with nontrivial commutation relations
\begin{equation}
[X^\mu,X^\nu]= i \lambda_P^2 Q^{\mu \nu},
\label{1}
\end{equation}
where the antisymmetric tensor $Q^{\mu \nu}$, depends, in general, on the
$X^\mu$. The noncommutative $*$-algebra ${\cal A}$, generated by regular
representations of Eq. (\ref{1}), contains all informations on what we may
call {\it
noncommutative spacetime}, while the geometric picture, which greatly helps
in the commutative cases, is lost.

The analogy with ordinary quantum mechanics
and quantization of phase space strongly suggests to those
who are fascinated by the path integral approach, as the author of
the present paper is, to establish a path
integral formulation of
these noncommutative geometries, taking the point of view that a
class of linear functionals
over ${\cal A}$, which turn into evaluation maps in the commutative
limit,
can be expressed as integrals, with a suitable measure, of ordinary
functions over the classical, commutative spacetime. The reason why such
an approach could eventually turn out to be useful is that it relates in
a rather simple way the algebra generated by the $X^\mu$ to its commutative
counterpart, and this is a powerful way to look for generalization at the
noncommutative level of the basic structures of Riemannian geometry: metric,
connections and curvature.

\section{Path integral over spacetime}

We start considering the pair $(M^{2n},\Omega)$, where $M^{2n}$ is an
even-dimensional
differentiable manifold and $\Omega$ a symplectic form, and introduce the
{\it generating functional}
\begin{equation}
Z(x_0,J) = N \int D[\gamma]~ \exp \left[ i \lambda_P^{-2}
\left( \int_{\Gamma} \Omega  +  (x,J) \right) \right],
\label{2}
\end{equation}
where $N$ is a normalization constant, $J$ an arbitrary source,
the integration is carried
over all closed curves $\gamma$ with base point $x_0 \in M^{2n}$
and $\Gamma$ is any
two--dimensional surface with boundary $\gamma$.
We have also defined
\begin{equation}
(x,J) = \int_\gamma x^\mu(\tau) J_\mu(\tau) d \tau,
\label{3}
\end{equation}
where $x^\mu(\tau)$ is any parameterization of $\gamma$.
We here illustrate the basic
construction for a manifold with trivial second homology group.
The general case will be discussed in section 3.
We therefore have
\begin{equation}
\int_{\Gamma} \Omega=
\int_\gamma A_\mu(x) {d \over d \tau}
x^\mu(\tau) d \tau,
\label{4}
\end{equation}
with $\Omega =d A$.
$Z(x_0,J)$ is invariant under curve
reparameterization, provided we redefine the
arbitrary external source $J$. In the following we will
consider $\tau \in [0,1]$.
Equation (\ref{2}) is clearly inspired by the
expression of path integral in phase space with external sources in the
coherent state representation.
The generating functional $Z(x_0,J)$ defines a
(noncommutative) algebra ${\cal A}$ generated by $2n$ elements
$X^\mu$,
implicitly defined via the introduction of a set of linear functionals
$\rho_{x_0}$ as follows
\begin{eqnarray}
& &\rho_{x_0}(X^{\mu_1}...X^{\mu_k})
\equiv \lim_{\tau_i-\tau_{i-1} \rightarrow 0^-}
N Z(x_0,0)^{-1}
\int D[\gamma]~ e^{i \int_{\Gamma} \Omega /\lambda_P^2}
x^{\mu_1}(\tau_1)...x^{\mu_k}(\tau_k) \nonumber \\
& & = \lim_{\tau_i-\tau_{i-1} \rightarrow 0^-}
\left. (- i \lambda_P^2)^k Z(x_0,0)^{-1}
{\delta \over \delta J_{\mu_1} (\tau_1)}...
{\delta \over \delta J_{\mu_k} (\tau_k)} Z(x_0,J) \right|_{J=0}.
\label{6}
\end{eqnarray}
The way the limit is taken guarantees the ordering of the $X^{\mu_i}$ in the
right-hand side of Eq. (\ref{6}). Once defined on all polynomials in the
$X^\mu$, Eq. (\ref{6}) can
be applied to the entire algebra of continuous functions in the weak
topology.

To understand this definition, we consider the {\it classical}
limit $\lambda_P \rightarrow 0$.
The integral over curves is then dominated by the
contribution
at stationary points, i.e.
$\Omega_{\mu \nu}(x) dx^\nu/d \tau =0$.
Since $\Omega$ is not degenerate, the leading term
satisfying $x^\mu(0)=x^\mu(1)=x_0^\mu$,
is therefore given by
the curve $x^\mu(\tau)=x_0^\mu,~\forall \tau$, and thus
\begin{equation}
\rho_{x_0}(X^{\mu_1}...X^{\mu_k})
\rightarrow \left.
x^{\mu_1} ...x^{\mu_k} \right|_{x_0} + {\cal O}(\lambda_P^2).
\label{7}
\end{equation}
The maps $\rho_{x_0}$ reduce to the evaluation maps ($*$-homomorphisms)
of the commutative
algebra of smooth functions over $M^{2n}$, $\rho_{x_0}: f(x) \mapsto f(x_0)$

The explicit evaluation of $Z(x_0,J)$ may be rather involved or even
impossible for a generic two form $\Omega$; in these cases
application of perturbation techniques,
as in ordinary path integral in quantum mechanics, may nevertheless
provide some information
on the underlying noncommutative spacetime.
In the simple case of a constant
$\Omega$, or assuming that it is a slowly varying function of $x$ on the
scale $\lambda_P$, one can easily compute Eq. (\ref{2}) finding \cite{b}
\begin{equation}
\rho_{x_0}(X^\mu)= x_0^\mu~~~,
\rho_{x_0}([X^\mu,X^\nu])= {i \over 2} \lambda_P^2
\Omega^{-1 \mu \nu} (x_0).
\label{8}
\end{equation}
As expected the commutator is proportional to the inverse symplectic form
$\Omega^{-1}$ and using
notation of Eq. (\ref{1}), we see that $X^{\mu}$ generates a
noncommutative algebra with
$\rho_{x_0}(Q^{\mu \nu})= \Omega^{-1 \mu \nu}(x_0)/2$.

\section{Non-trivial topologies and Black Hole area quantization}

In the case of a manifold
$M^{2n}$ with nontrivial second homology group
the measure in the integral over closed curves in Eq. (\ref{2}) is,
in general, a multivalued function, since the integral of $\Omega$
over two surfaces $\Gamma$
and $\Gamma'$, both with boundary $\gamma$, may be different if
$\Gamma-\Gamma'$ is a non trivial 2-cycle. The ambiguity in the choice of the
surface $\Gamma$, however, can be removed by requiring {\it quantization}
conditions
on the integral of $\Omega$ over a set of generators $\Gamma_i$
of $H^2(M^{2n})$
such that $exp~(i \int_{\Gamma_i} \Omega \lambda_P^{-2}$) is
single valued, analogous
to the one introduced to quantize the motion of a
particle in the field of a magnetic charge.

Let us consider for example the pair $(S^2,\Omega)$. The generalization to
an arbitrary manifold is straightforward \cite{b}.
For any closed curve
the measure in the path integral will
be always single valued if we require
that the integral of $\Omega$ over $S^2$ is a multiple of $2 \pi$, in unit
$\lambda_P^2$
\begin{equation}
\int_{S^2} \Omega = 2 n \pi \lambda_P^2
\label{9}
\end{equation}
We stress the point that requiring Eq. (\ref{9}) is a consistency condition
to construct, via the path integral approach outlined in this paper, a
deformation of spacetime, and so it seems to us intimately
related to the appearance of microscopic noncommutativity at the Planck
scale.
Actually, we observe that
what we have just discussed could have an intriguing relationship with the
idea that
black hole horizon area is quantized and its spectrum is uniformly spaced
\cite{c}. This is in fact what condition (\ref{9}) is stating, if $S^2$
represents the black hole horizon area and with $\Omega$ the area two-form.
Black hole physics is probably one of the best scenarios where the
structure of spacetime at small scales, otherwise unobservable,
could be felt well above the Planck length. The topological constraint
(\ref{9}) may be, perhaps, a clue in the direction of a deep interplay
of microscopic noncommutativity and macroscopic phenomena such as black hole
mass quantization.

\end{document}